\documentclass[smallabstract,smallcaptions]{dccpaper}
\pdfoutput=1
\usepackage{epsfig, bm, booktabs, multirow}
\usepackage{citesort}
\usepackage{amsmath}
\usepackage{amssymb}
\usepackage{color}
\usepackage{url}

\newlength{\figurewidth}
\newlength{\smallfigurewidth}

\newcommand{\etal}{\textit{et al. }}

\newcommand{\e}[1]{\!\times\!10^{#1}}

\newcommand{\image}{x}
\newcommand{\reco}{\hat{x}}
\newcommand{\mask}{m}

\newcommand{\lambdat}{\lambda_\mathrm{t}}
\newcommand{\lambdae}{\lambda_\mathrm{e}}

\newcommand{\enc}{g_\mathrm{a}}
\newcommand{\dec}{g_\mathrm{s}}
\newcommand{\params}{{\bm{\theta}}}

\begin{document}

\title
{\large
\textbf{Learning True Rate-Distortion-Optimization for End-To-End Image Compression}
}

\author{%
Fabian Brand, Kristian Fischer, Alexander Kopte, and Andr\'e Kaup\\[0.5em]
{\small\begin{minipage}{\linewidth}\begin{center}
\begin{tabular}{c}
Multimedia Communications and Signal Processing  \\
Friedrich-Alexander University Erlangen-Nuremberg\\
Cauerstr. 7, 91052 Erlangen, Germany\\
\url{{fabian.brand, kristian.fischer, alex.kopte, andre.kaup}@fau.de}
\end{tabular}
\end{center}\end{minipage}}
}

\maketitle
\thispagestyle{empty}

\begin{abstract}
Even though rate-distortion optimization is a crucial part of traditional image and video compression, not many approaches exist which transfer this concept to end-to-end-trained image compression. Most frameworks contain static compression and decompression models which are fixed after training, so efficient rate-distortion optimization is not possible. In a previous work, we proposed RDONet, which enables an RDO approach comparable to adaptive block partitioning in HEVC. In this paper, we enhance the training by introducing low-complexity estimations of the RDO result into the training. Additionally, we propose fast and very fast RDO inference modes. With our novel training method, we achieve average rate savings of 19.6\% in MS-SSIM over the previous RDONet model, which equals rate savings of 27.3\% over a comparable conventional deep image coder.
\end{abstract}

\Section{Introduction}
Learning-based methods for image and video compression have received broad attention in research over the last years. As in traditional methods like JPEG~\cite{Wallace1992_JPEGstillpicture}, JPEG2000~\cite{ITUTI2004_JPEG2000Image}, HEVC~\cite{SullivanOH2012_OverviewHighEfficiency}, VVC~\cite{BrossCL2020_VersatileVideoCoding} or AV1~\cite{HanLM2021_TechnicalOverviewAV1}, these methods rely on a transformation of an image into a sparse domain. In contrast, in learning-based approaches, the transformation is not a handcrafted linear transform such as DCT, but rather a data-driven non-linear transform. 

Another major difference to most traditional methods is the processing of the image as a whole or in large tiles, whereas for example JPEG or HEVC process the image in a block-wise fashion. The block-wise processing grants several advantages but also disadvantages in the coding process. The block structure enables spatial prediction using previously decoded blocks as reference. Adaptive block partitioning can be used to allocate the rate optimally in the picture and can be used to generate blocks with good spatial correlation when object and block boundaries coincide. On the other hand, block partitioning produces block artifacts which require special deblocking filters for concealment.

Especially the possibility of adaptive block-partitioning is of great advantage. The recent development of VVC, where compared to HEVC also rectangular blocks were possible, showed that large portions of the gains were due to this extended block partitioning possibilities~\cite{WieckowskiMG2019_Generalizedbinarysplits}. The encoder can pick from even more flexible structures and reach optimal points. In praxis, the encoder tests multiple configurations and chooses the best. This is part of the rate-distortion-optimization (RDO), which is greatly beneficial for video compression.

In deep-learning-based compression algorithms, such as \cite{BalleLS2017_Endendoptimized,BalleMS2018_Variationalimagecompression,MinnenBT2018_JointAutoregressiveHierarchical,ChengST2020_LearnedImageCompression}, RDO is performed during training. The compressive autoencoder which is the core of the methods is trained to minimize a joint rate-distortion loss-function
\begin{equation}
	L = D+\lambda R,
\end{equation}
with distortion $D$ and rate $R$. The parameter $\lambda$ determines the location of the resulting rate point on the Pareto front of this multidimensional optimization problem. The network parameters, which were trained using this loss function, only depend on the training data and can not be adapted to the current image to be coded. That way, one can interpret this method as a \emph{static} RDO. In contrast, the adaptive block partitioning in traditional coding is performed during the encoding and depends on the image, and thus can be seen as a \emph{dynamic} RDO.

In this work, we build on RDONet~\cite{BrandFK2021_RateDistortionOptimized}, a recently proposed hierarchical autoencoder structure, which enables a dynamic RDO similar to adaptive block partitioning. In this paper, we boost the performance of RDONet with a novel training procedure. Using heuristic methods, we can approximate the RDO result and generate more realistic training conditions. We thereby fit the network better to the final task. Additionally, we propose a fast encoding mode based on the same heuristics.

\Section{Related Work}
One of the first works in end-to-end trained image compression was written by Ball\'e \etal in 2017~\cite{BalleLS2017_Endendoptimized}. There, the authors proposed a framework using an autoencoder with an entropy bottleneck. In subsequent publications, this framework was extended in order to provide better probability models to compress the latent space. In \cite{BalleMS2018_Variationalimagecompression}, a hyperprior network was proposed which transmits the variance of each latent space entry on a separate channel. This concept was extended in \cite{MinnenBT2018_JointAutoregressiveHierarchical}, which introduced an autoregressive context model to further enhance the probability models for the latent space.

In further developments, \cite{ChengST2020_LearnedImageCompression}, the authors use a model which uses Gaussian mixture models (GMMs) to model the latent space distribution instead of a single Gaussian or Laplacian distribution, further improving the compression of the latent space. Furthermore, this model makes use of attention layers and residual blocks in both encoder and decoder, thus increasing the capacity of the network. In \cite{LiAL2020_DeepLearningBased}, the authors propose to use trellis coded quantization instead of scalar quantizer.

The parameters of the mentioned networks are all fixed after training and both encoding and decoding function can not be changed during coding. In \cite{SchaferPP2021_RateDistortionOptimization}, Sch\"afer \etal proposed a rate distortion optimization algorithm, which optimizes the quantized coefficients of the latent space. The authors also proposed a fast search algorithm to speed up the cumbersome full RDO.

\Section{RDONet}
The compression network RDONet is inspired by traditional block-based image and video compression, as for example in HEVC. There, the image is compressed on different granularities depending on the image characteristics. For stationary image content, large blocks are used, so large-scale spatial redundancy can be efficiently exploited, while detailed areas are coded in smaller blocks. Thereby, more bits are spent to compress the more complicated content adequately. 

End-to-end image compression is typically performed on full images or at least on very large areas at once, since one element in the latent space has a very large field of view and influences large areas in the image. However, multiple studies e.g. in \cite{BalleMS2018_Variationalimagecompression} or \cite{ChengST2020_LearnedImageCompression} have shown that the variance of the latent space remains larger in areas with many details and lower in areas with flat regions. Those flat areas could benefit from further increasing the field of view to capture the spatial correlations on an even wider scale. Wider spatial context can be achieved by a deeper autoencoder. RDONet therefore is designed that different parts of the image can be coded at different depth. Similar to adaptive block partitioning in traditional video compression, the depth for each area shall be determined at encoding time. 

\begin{figure}
	\vspace{-0.5cm}
	\footnotesize
	\begin{tabular}{ccc}
		\includegraphics[width=0.28\linewidth]{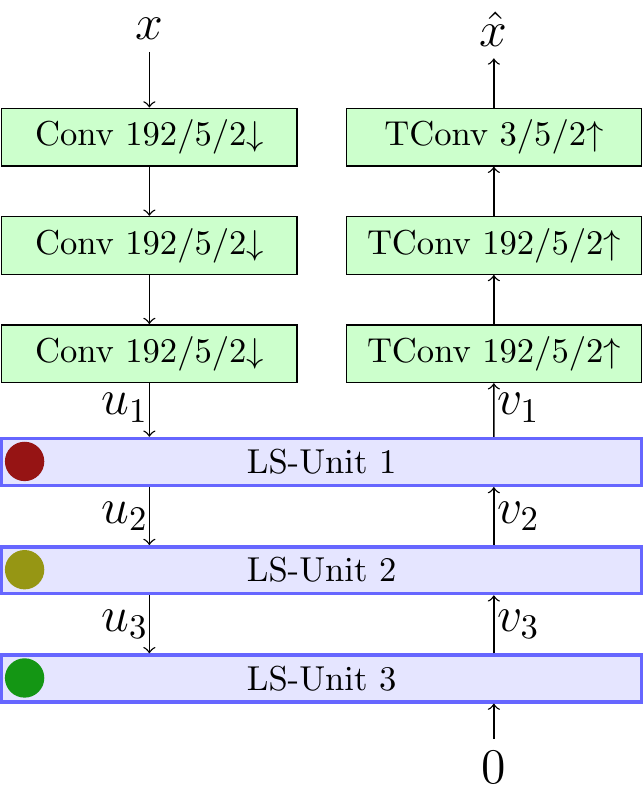}&
		\includegraphics[width=0.37\linewidth]{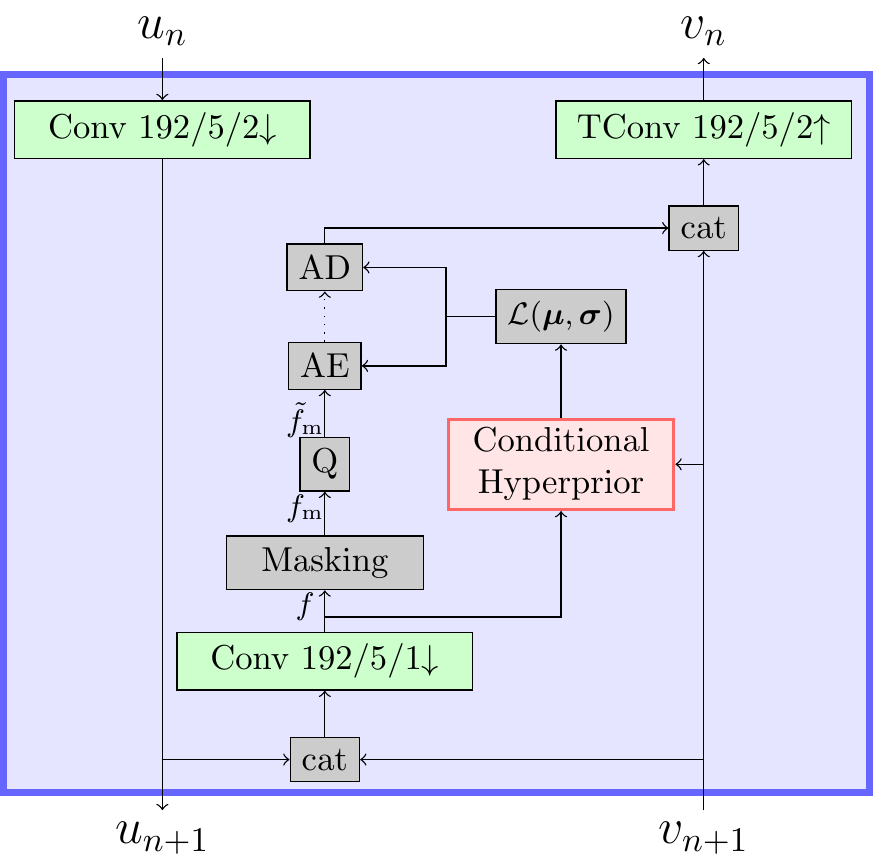}&
		\includegraphics[width=0.28\linewidth]{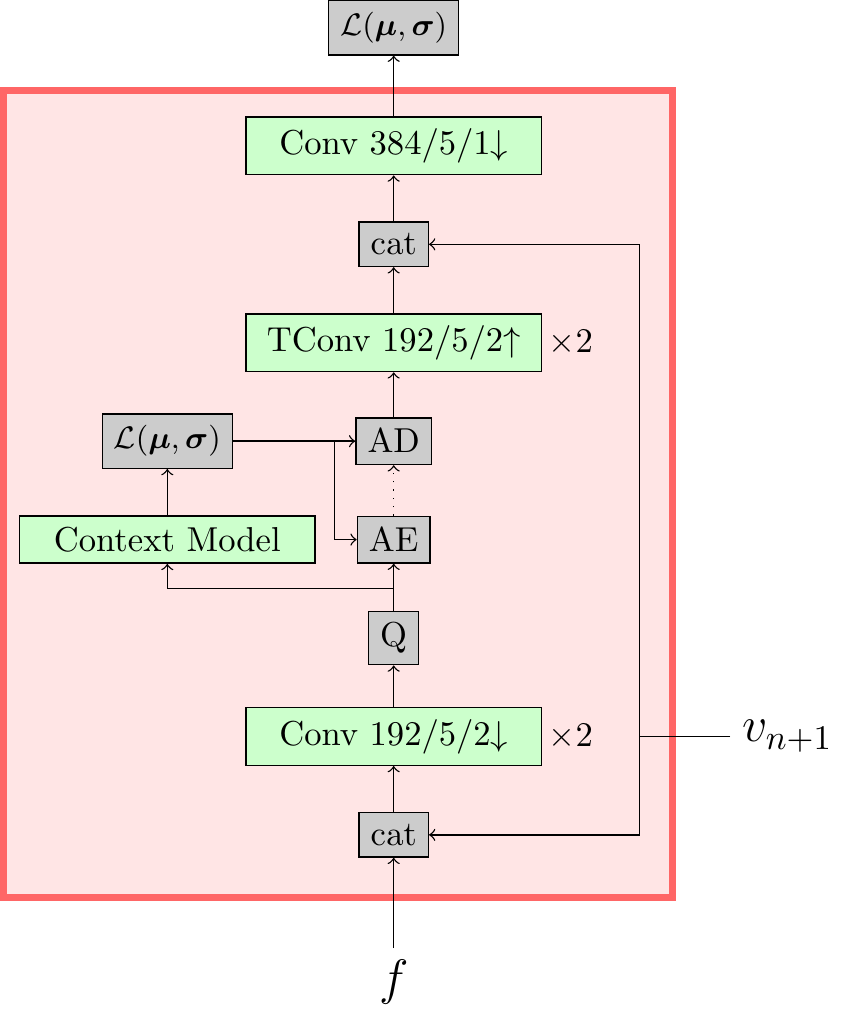}\\
		(a) RDONet & (b) LS-Unit & (c) Conditional Hyperprior\\
	\end{tabular}
\caption{Network structure of RDONet. Conv $c/k/s\downarrow$ and TConv $c/k/s\uparrow$ denote convolutional and transposed convolutional layers with $c$ output channels, $k\!\times\!k$ kernels and a stride of $s$. All downsampling or upsampling convolutions in (a) and (b) are followed by a (inverse) generalized divisive normalization (GDN/IGDN) layer~\cite{BalleLS2015_DensityModelingImages}. In the hyperprior network (c), PReLUs are used. $\image$ and $\reco$ denote the orignal and reconstructed image, respectively. $u_n$ and $v_n$ denote the representations at each latent space and $f$ denotes the latent representation to be transmitted before masking. AD, AE and Q are arithmetic encoder and decoder, and quantizer respectively. $\mathcal{L}$ denotes a Laplace distribution. The colored dots in the LS-Units in (a) are used in Figs. \ref{Fig:ExampleSigma}, \ref{Fig:Examples} and \ref{Fig:RD} for reference.} \label{Fig:Schematics}
\vspace{-0.5cm}
\end{figure}

RDONet contains three separate latent spaces at different depths. Together with a mask which has to be transmitted as side information, this allows us to spatially control the compression granularity. The masked elements are set to zero in the masking layer. They are not explicitly transmitted, since the decoder receives the mask and can therefore reconstruct the zeros. The depth of the latent space is analogous to the block sizes in classical coders. A deeper latent space corresponds to a larger block size. Note that one pixel in the latent space influences a much larger area than just the corresponding block in the image. Therefore, the blocks can not be optimized independently as we will show later on.

We show the full structure of RDONet in Fig.~\ref{Fig:Schematics}. We see that the network contains three downsampling convolutional layers to generate an intermediate feature representation before the first latent space. The three latent spaces are processed in so-called latent space units (LS-Units). In each LS-Unit another downsampling convolution is performed before the latent representation is masked and transmitted. Transmitting three latent spaces instead of one may lead to worse performance, if all latent spaces are treated as independent. In fact, however, knowledge of one latent representation may enhance the coding of the other. 
To overcome this challenge, we predict each latent representation with the previously transmitted representation of a lower resolution and employ a conditional hyperprior. After decoding the latent space, the previous and the current representation are concatenated and jointly upsampled with a transposed convolution. 

We want to point out here, that the terms ``split'' and ``partitioning'' are mainly used in analogy to adaptive block-partitioning. This however does not mean that the image is actually split and coded separately or that block artifacts appear in the resulting image.

\Section{Enhanced Training Procedure}
On a high level, RDONet consists of two functions $\enc\!\left(\image|\mask,\params\right)$ and $\dec\!\left(y|\mask,\params\right)$, which are the analysis and synthesis function, respectively. Both functions depend on the network parameters $\params$ and---different to conventional compression networks---also on a particular mask $\mask$. Here, $\image$ denotes the original image and $y$ denotes the collection of latent representations. The network parameters $\params$ are determined during training and are the solution of the following minimization problem:
\begin{equation}
	\params = \underset{\params}{\operatorname{argmin}} ~ D\left[\,\dec\!\left(\,\enc\!\left(\image|\mask,\params\right)|\mask,\params\right), x\right] + \lambda R\left[\,\enc\!\left(\image|\mask,\params\right)\right].
\end{equation} 
Since the optimum depends on the chosen mask, the mask has also to be set during training.  

In \cite{BrandFK2021_RateDistortionOptimized}, randomly generated masks were used during training. These masks were independent of the input image, which does not optimally reflect the use-case, as during inference the mask is determined to minimize the RD loss function. To take this into account during training, we have to find (near) optimal masks during training time. However, these masks also depend on the current network parameters.

To solve this chicken-egg problem several solutions come to mind. In general, an iterative approach seems possible. Here first the masks are optimized for a given network, which is then trained with those masks. However, computing the optimal masks using RDO is computationally expensive, since the network has to be run multiple times. That way, training with this approach is currently infeasible.

Another conceivable method would be to use a neural network to estimate the optimal mask and train both networks jointly. However, the masking layer performs a (hard) masking of the input, which is a discrete operation. Therefore, the masking layer is not differentiable with respect to the mask. 

Since both the large time consumption and the non-differentiability of the masking layer inhibit a training with optimal masks, we propose to estimate the mask using traditional signal processing methods. In particular, we propose a variance criterion to compute the masks based on the image content. This strategy is inspired by observations of block partitioning in hybrid video coders. There, areas with larger variance are often encoded with smaller block size, since those areas contain more details. Areas with smaller variance are often coded in larger blocks, as they usually contain more planar content. Of course, this is only a rough estimation of the block partitioning as for example in HEVC. However, our goal is not to train a coder which exactly takes over HEVC block partitioning, but rather to train the coder to get a better sense of the content which is expected in each latent space level.

\begin{figure}
	\centering
	\footnotesize
	\vspace{-0.5cm}
	\begin{tabular}{cc}
		\includegraphics[width=0.44\linewidth]{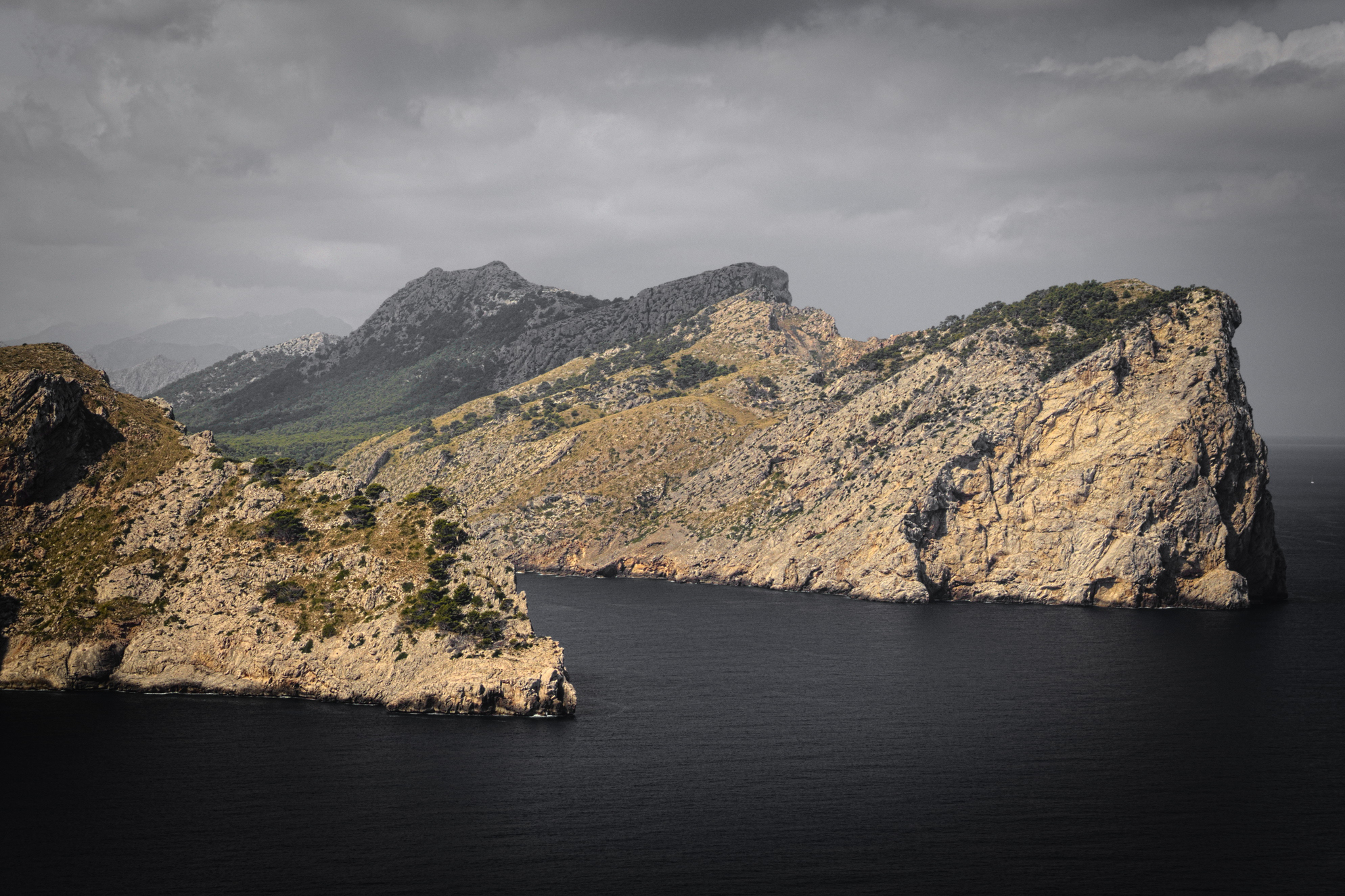}&
		\includegraphics[width=0.44\linewidth]{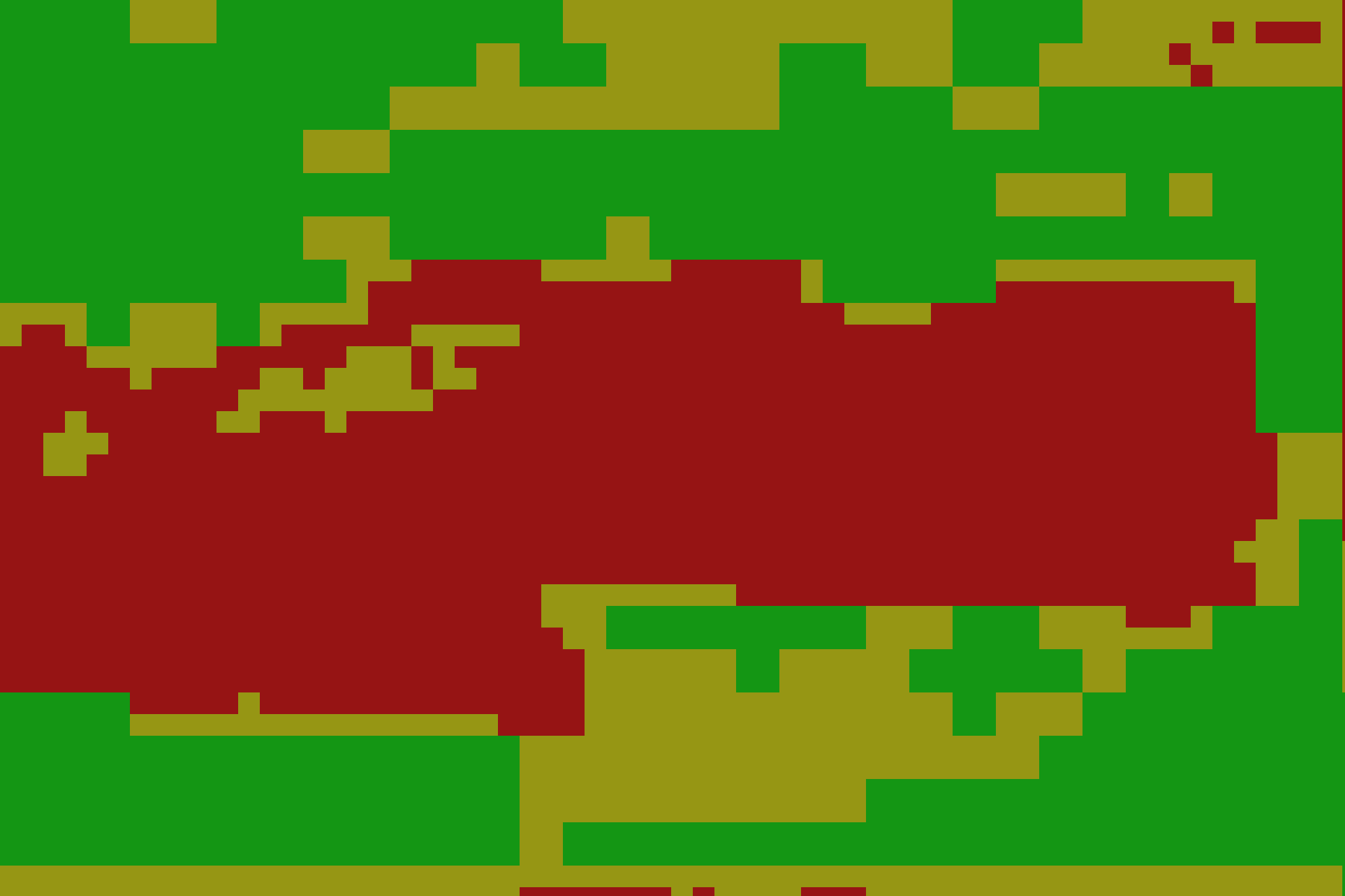}\\
		(a) Image& (b) Blocks generated with variance criterion\\
	\end{tabular}
	\caption{(a) Image \textit{wojciech-szaturski-3611} from the CLIC validation set. (b) Blocks generated with the variance criterion. Red areas are to be coded finely in the highest latent space and green areas are coded coarsely in the lowest latent space. Yellow areas are to be coded in the medium latent space. The color mapping is also indicated in Fig. \ref{Fig:Schematics} with the red, yellow and green dots in the respective latent space units.}\label{Fig:ExampleSigma}
	\vspace{-0.3cm}
\end{figure}

We propose the following mask generation process. First, we split the image in blocks of size $64\!\times\!64$, which is the largest possible block size. For each block, we compute the variance of all pixel values and sum them up over all three color channels. If this variance is larger than a certain threshold $\sigma^2_\mathrm{th,1}$, then the block is split. For all sub-blocks, we perform the same procedure, splitting the block if a variance of $\sigma^2_\mathrm{th,2}$ is exceeded. With this simple procedure, we are able to efficiently split the image sensibly. Differently to the random initialization from \cite{BrandFK2021_RateDistortionOptimized}, the content which is encoded by each latent space level shares some characteristics, so the latent space can specialize on that characteristic. Preliminary experiments have shown that the precise values of $\sigma^2_\mathrm{th,1}$ and $\sigma^2_\mathrm{th,2}$ are not actually critical for the performance of the network. The rough classification in patches of low, medium, and large variance seems to be sufficient. Minor deviations are compensated by the training approach. In Fig. \ref{Fig:ExampleSigma}, we show the masks of one example image. We chose $\sigma^2_\mathrm{th,1}=5\e{-4}$ and $\sigma^2_\mathrm{th,2}=2\e{-3}$. The result roughly meets the expectations. The sky and sea are mostly coded in the lowest and middle latent space (green and yellow), while the rock formations are coded in the highest latent space with higher spatial resolution. These masks are now used as input to the training to approximate the true RDO in inference.

\Section{Fast RDO Search}
After we obtain a trained network, we have to perform a rate-distortion optimization to find the optimal encoding configuration for each image. Here, we first follow the approach from \cite{BrandFK2021_RateDistortionOptimized}. 
In this approach, the image is first divided in blocks of size $64\!\times\!64$. Each block is initialized to be coded with the coarsest latent space. Afterwards, each block is optimized individually. Since the decisions also heavily influence the performance of the neighboring blocks, no parallel processing is possible. At the time when one block is optimized, many surrounding blocks still have their initial, non-optimal configuration. Therefore, this process does not result to a global optimum. To counter this effect a two-pass RDO was proposed in \cite{BrandFK2021_RateDistortionOptimized}. The second RDO pass optimizes the blocks again, but this time when all surrounding blocks are initialized with the result of the first RDO pass. It was shown that this results in superior coding performance and is thus closer to the global optimum. With the algorithm from \cite{BrandFK2021_RateDistortionOptimized}, the network has to be run 6 times per pass and block. Depending on the image size, this procedure may take a long time.

We therefore propose to rely on the same heuristics which we used to improve our training to find a new initialization for the RDO. By using the same variance criterion, we can determine a rough estimation of the optimal partitioning. That way, we can introduce a fast mode, which uses the new initialization as starting point for the RDO. We argue that the result from the variance-based splitting is good enough that the RDO converges to the optimum in only one pass. That way, we can halve the complexity of the RDO and speed up encoding by 50\%.

Furthermore, we propose a very fast mode for RDONet, which does not perform the RDO anymore, but instead uses the variance-based masks directly. That way, the encoding time is further reduced, since we only need one pass through the network for the encoding. This variant requires at most 5 variance computations per $64\!\times\!64$ block, instead of several runs of the full network. 

\Section{Experiments}
\SubSection{Setup}
We implemented our networks using the PyTorch framework. To train the network, we chose a combined MSE and MS-SSIM loss function:
\begin{equation}
L_\mathrm{train} = D_\mathrm{MSE}(\image, \reco) + 0.1\cdot D_\mathrm{MS-SSIM}(\image,\reco) + \lambdat \cdot r,
\end{equation} 
where $\image$ and $\reco$ are the original and reconstructed image, respectively. The multiplier $\lambdat$ controls the trained rate point, and $r$ denotes the estimated rate over all latent spaces. Furthermore, $D_\mathrm{MSE}(\cdot,\cdot)$ is the MSE between two signal, and $D_\mathrm{MS-SSIM}(\cdot,\cdot)$ is the MS-SSIM loss.
We chose this combined loss function, since preliminary tests have shown that the network does not initially converge well when trained only with MS-SSIM.

We train all networks on a combination of the training set of the CLIC intra coding challenge 2021, the TECNICK dataset \cite{AsuniG2013_TESTIMAGESLargeData}, and the DIV2K dataset \cite{AgustssonT2017_NTIRE2017Challenge}. This accounts for a total of 1585 images with a variety of characteristics and resolutions. For training, we use the ADAM optimizer~\cite{KingmaB2015_AdamMethodStochastic} with an initial learning rate of 0.00001, which we divide by ten after the first 1000 epochs. We use patches of size $512\!\times\!512$ and a batch size of 8.

In preliminary experiments, we found that it is beneficial to train the network first with randomly generated masks, as in \cite{BrandFK2021_RateDistortionOptimized}, for 2000 epochs and then fine-tune the network with adaptive masks for another 600 epochs. As in the shown example in Fig. \ref{Fig:ExampleSigma}, we choose $\sigma^2_\mathrm{th,1}=5\e{-4}$ and $\sigma^2_\mathrm{th,2}=2\e{-3}$. All models are trained for four different values of $\lambdat\in\left\{0.01,0.02,0.04,0.08\right\}$. 

We test our framework on the validation set of the CLIC intra coding challenge 2021. The set consists of 41 high quality images which were collected from Unsplash. We compare our new model (denoted as RDONet-Var) against the model from \cite{BrandFK2021_RateDistortionOptimized} and a standard compressive autoencoder without hierarchical latent spaces. The reference network has a hyperprior network and a context model and is therefore comparable to the network proposed in \cite{MinnenBT2018_JointAutoregressiveHierarchical}. Otherwise we followed the same design decisions as RDONet to have maximal comparability.

For our experiments, we perform an RDO to minimize the loss function:
\begin{equation}
	L_\mathrm{RDO} = D_\mathrm{MS-SSIM}(\image,\reco) + \lambdae \cdot r,
\end{equation}
where $\lambdae$ is a freely selectable parameter controlling the trade-off between rate and distortion during inference. Depending on the choice of $\lambdae$, we obtain different rate points. By picking different $\lambdae$, we can achieve a certain flexibility in rate with only one trained model, which is not taken for granted in deep-learning-based compression. To compare ourselves against the base model, we assign a fix $\lambdae$ to every value of $\lambdat$ to obtain a rate-distortion curve with four points. Analogously to \cite{BrandFK2021_RateDistortionOptimized}, we choose to assign the values $\lambdae = [0.5,0.5,0.25,0.125]$ to $\lambdat = [0.08, 0.04, 0.02, 0.01]$. Note that we assign larger values of $\lambdae$ to larger values of $\lambdat$. This fits the intuition that both values should match. Since the loss functions in RDO and training differ, the multipliers also assume different values.

\begin{figure}
	\footnotesize
	\hspace{-0.3cm}\begin{tabular}{cccc} 
		\includegraphics[width=0.23\columnwidth] {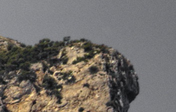}&
		\includegraphics[width=0.23\columnwidth] {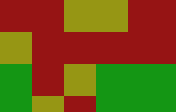}& 
		\includegraphics[width=0.23\columnwidth] {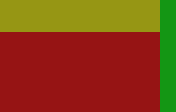}&
		\includegraphics[width=0.23\columnwidth] {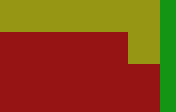}\\
		\includegraphics[width=0.23\columnwidth] {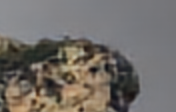} &
		\includegraphics[width=0.23\columnwidth] {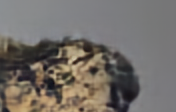} & 
		\includegraphics[width=0.23\columnwidth] {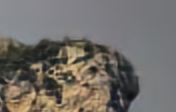} & 
		\includegraphics[width=0.23\columnwidth] {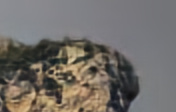}\\
		(a) Baseline &(b) RDONet\cite{BrandFK2021_RateDistortionOptimized} &(c) Fast Mode &(d) Very Fast Mode \\
		25.31\,dB/0.14bpp & 25.86\,dB/0.16bpp & 26.95\,dB/0.14bpp & 26.93\,dB/0.14bpp
		
	\end{tabular}
\caption{Zoomed-in examples of the image \textit{wojciech-szaturski-3611} when coded with different methods. In the top left corner we show the original image. The remainder of the top row shows the masks in the color scheme indicated in Fig. \ref{Fig:Schematics}.}\label{Fig:Examples}
\vspace{0.2cm}
\end{figure}

\SubSection{Results}
At first, we want to show the superior performance of RDONet-Var compared to the non-hierarchical model and the basic RDONet. We show the image \textit{wojciech-szaturski-3611} in Fig. \ref{Fig:Examples}. All images were coded with around 0.15\,bpp. Note that the local bit-rate may differ. In Fig. \ref{Fig:Examples}(a), we see the result of the non-hierarchical baseline coder, which exhibits blurry behavior in the highly structured rock formations. When we look at the basic RDONet after a two-pass RDO in (b), we see that the rocks are less blurry. However, when we look at the corresponding masks, we see a rather erratic behavior. This changes, when we switch to RDONet-Var in (c) and (d). In the fast RDO mode, the rocks are now all coded in the fine latent-space and the sky in the lowest latent space. With the new training, the network gets better at focusing on structured areas. Since more bits are allocated in the rocky area, we achieve a much less blurry image and preserve more details. When we switch to the very fast mode, we see that both mask and image do not change much. This already indicates the good performance of the very fast mode.

\begin{figure}
	\footnotesize
	\begin{tabular}[b]{cc}
		
		\begin{tabular}[b]{cc}
			\includegraphics[width=0.22\columnwidth] {Images/wojciech-szaturski-3611.png}
			&\includegraphics[width=0.22\linewidth]{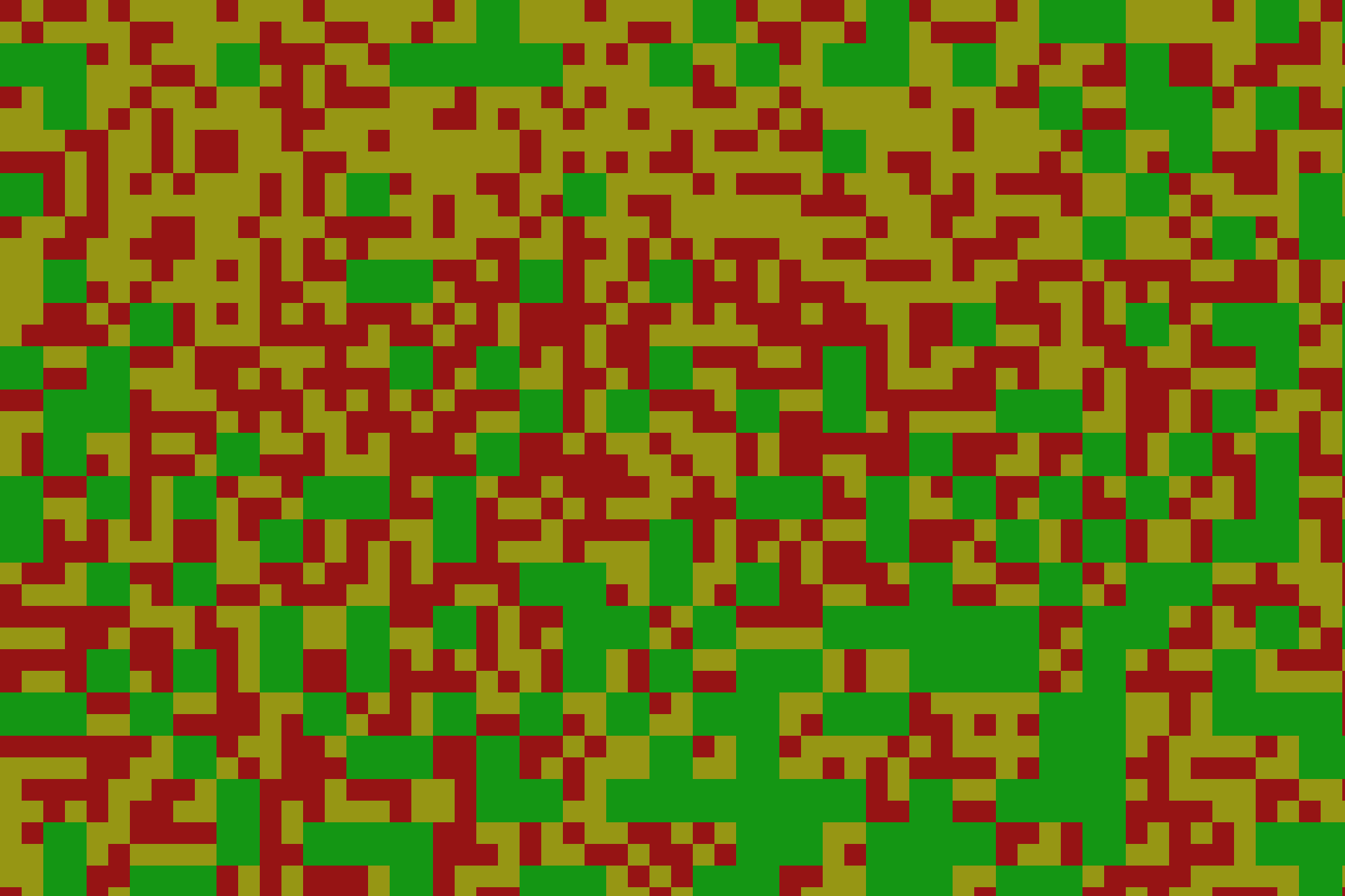} \\
			(a) Image & (b) Basic RDONet \cite{BrandFK2021_RateDistortionOptimized}\\
			\includegraphics[width=0.22\linewidth]{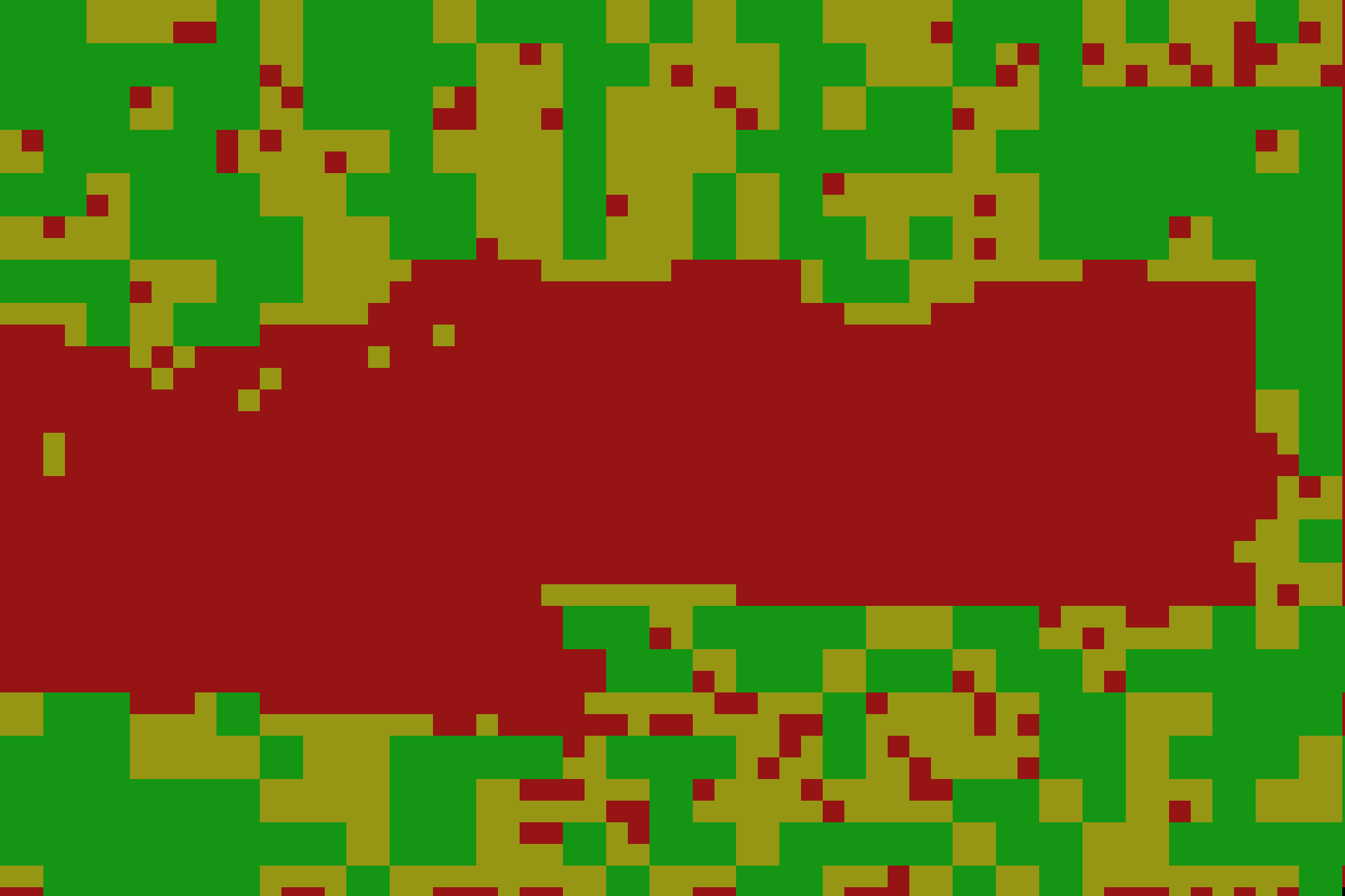}
			&\includegraphics[width=0.22\linewidth]{Images/wojciech-szaturski-3611_p0_m0_r1_Init_masknc.png}\\
			(c) Fast Mode& (d) Very Fast Mode
		\end{tabular}
		
		&\begin{tabular}[b]{c} \includegraphics[width=0.40\linewidth]{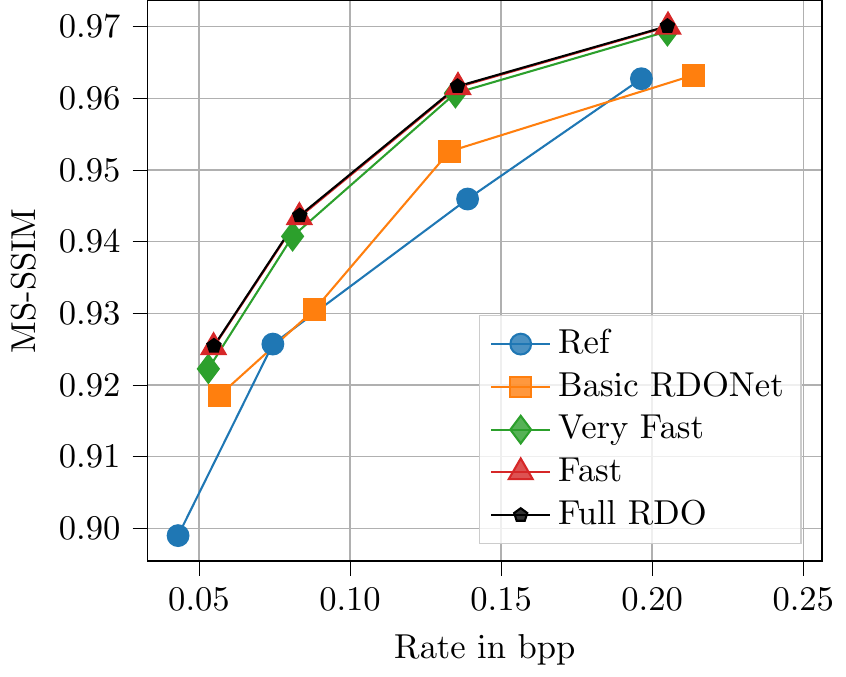} \\ (e) Rate distortion curve \end{tabular}\\
	\end{tabular}
	\caption{(a)-(d) Image and full masks of \textit{wojciech-szaturski-3611}. (e) Rate-distortion diagram of \textit{wojciech-szaturski-3611}.}\label{Fig:RD}
	\vspace{-0.4cm}
\end{figure}

In Fig. \ref{Fig:RD} (b)-(d), we show the full masks of the example image. This confirms that the masks obtained with the new model are of higher quality. The mask of the basic RDONet has the correct tendencies but overall behaves erratically. In \ref{Fig:RD} (e), we show the rate-distortion diagram for this image. We see that the new RDONet outperforms the old model significantly and for all measured rate points. However, all variations of RDONet perform better than the non-hierarchical model. Furthermore, we see that the fast mode performs better than the very fast mode. However, this difference is quite small. In most practical cases the very fast encoding mode is the better choice since no cumbersome RDO search is necessary. We also see that the difference between the fast one-pass RDO and the full two-pass RDO is marginal for this image. This again shows the high quality of the novel initialization technique. 

\begin{table}
	\centering
	\small
	\begin{tabular}{l|cc|cc|ccc}
		\toprule
		Network & \multicolumn{2}{c|}{Basic RDONet \cite{BrandFK2021_RateDistortionOptimized}} & \multicolumn{5}{c}{RDONet-Var (Ours)} \\
		RDO-Init &  \multicolumn{2}{c|}{Static} & \multicolumn{2}{c|}{Static} &  \multicolumn{3}{c}{Variance Adaptive} \\
		RDO-Passes & 1 & 2 & 1 & 2 & 0 & 1 & 2 \\

		\midrule
		Best Case & -18.9\% & -22.5\% & -43.7\%& -45.3\%& -36.6\% & -44.4\% & -45.2\%\\
		Worst Case & +7.5\% & +3.5\% & -11.9\% & -12.5\% & -6.67\% & -11.9\% & -12.5\% \\
		Average & -4.1\% & -7.7\% & -23.3\% & -25.0\% & -23.6\% & -26.8\% & - 27.3\% \\
		\midrule
		RDO-Complexity & $6\cdot N_{64}$ & $12\cdot N_{64}$ & $6\cdot N_{64}$ & $12\cdot N_{64}$ & 0 & $6\cdot N_{64}$ & $12\cdot N_{64}$ \\
		\bottomrule
	\end{tabular}
\vspace{-0.1cm}
\caption{BD-Rate savings over the entire CLIC validation dataset. Negative values denote rate savings. We compare the Basic RDONet from \cite{BrandFK2021_RateDistortionOptimized} with our proposed model RDONet-Var. The RDO complexity is measured in number of necessary network runs per image, where $N_{64}$ is the number of $64\!\times\!64$ blocks in that image.}\label{Tab:BDR}
\vspace{-0.4cm}
\end{table}

In Tab. \ref{Tab:BDR}, we show the Bj\o ntegaard Delta rate (BD-Rate) \cite{Bjontegaard2001_CalculationaveragePSNR} averaged over all images. We compare the basic RDONet with RDONet-Var. At first, we look at a static RDO initialization with the lowest layer, as proposed in \cite{BrandFK2021_RateDistortionOptimized}. After two passes, the old RDONet achieves already 7.7\% rate savings over the baseline. However, when we perform our enhanced training, we can increase the savings to 25.0\%. Also in the best case and worst case images, the performance increases. It is notable, that we save at least 12.5\% rate, even for the worst performing image.

To use our fast mode, we initialize the RDO with the variance-based masks. When we perform a full RDO with these initial values, we can increase the rate savings to 27.3\%. Even with the fast mode  we can achieve 26.8\% rate savings, which is more than the full RDO with static initialization. Note that the effect of the second pass reduces with initialization. With static initialization, we save additional 1.7\% with the second pass. With adaptive initialization this gap reduces to 0.5\%, suggesting that the result is already close to the optimum.

With the very fast mode, we achieve on average 23.6\% rate savings. We see that the RDO steps can further improve the performance further, but the very fast mode, which only estimates the optimal configuration, already performs very good. As indicated in the table, this method does not require any additional coding passes. 

\Section{Conclusion}
In this paper, we achieved a considerable improvement in rate-distortion-optimized deep image compression. By enhancing the training of RDONet using special content-dependent masks, we were able to increase the average rate savings over a conventional deep image coder from 7.7\% to over 25\%. Furthermore, we proposed strategies to reduce the complexity of the RDO. By initializing the configuration with estimated masks, we can improve the result further and speed up the RDO at the same time. We find that even the estimation can achieve large gains. This variant does not require any RDO search.

In future work, this approach can be extended to video compression, where adaptive block partitioning is firmly established in traditional coders. We therefore expect similar gain in this field, in particular we can thereby take the spatially varying inter prediction quality into account.
Additionally, we aim to extend the RDO to account for the human visual system by introducing perceptually motivated loss functions. The principles used in this work are sufficiently general, so training with any loss function is possible

\Section{References}
\bibliographystyle{IEEEbib}

\end{document}